\documentclass[12pt]{article}
\usepackage{amsmath,amssymb,amsfonts,graphicx,verbatim,cite}
\usepackage{wrapfig}
\pdfoutput=1

\usepackage[usenames,dvipsnames]{xcolor}

\usepackage{hyperref}
\hypersetup{colorlinks=true, linkcolor=BrickRed, citecolor=Violet, filecolor=OliveGreen, urlcolor=RoyalBlue, filebordercolor={.8 .8 1}, urlbordercolor={.8 .8 0}}

\definecolor{palatinatepurple}{rgb}{0.41, 0.16, 0.38}

\definecolor{uglybrown}{rgb}{0.8,  0.7,  0.5}
\definecolor{darkgreen}{rgb}{0,.5,0}

\renewcommand{\title}[1]{\vbox{\center\LARGE{#1}}\vspace{5mm}}
\renewcommand{\author}[1]{\vbox{\center#1}\vspace{5mm}}
\newcommand{\address}[1]{\vbox{\center\em#1}}

\renewcommand{\title}[1]{\vbox{\center\bf{\Large{#1}}}\vspace{5mm}}

\numberwithin{equation}{section}

\newcommand{\beq}{\begin{equation}}
\newcommand{\eeq}{\end{equation}}
\newcommand{\be}{\begin{equation}}
\newcommand{\ee}{\end{equation}}

\def\bea{\begin{eqnarray}}
\def\eea{\end{eqnarray}}

\def\({\left(}
\def\){\right)}

\def\CM{{\cal M}}

\def\CW{{\cal W}}

\newcommand{\lads}{\ell_{\text{AdS}}}

\begin{document}

\begin{titlepage}

\hfill \\
\hfill \\
\vskip 1cm

\title{Complexity Equals Action}

\author{Adam R. Brown${}^a$, Daniel A. Roberts${}^b$, Leonard Susskind${}^a$,  \\ Brian Swingle${}^a$, and Ying Zhao${}^a$ }

\address{${}^a$Stanford Institute for Theoretical Physics and\\
Department of Physics, Stanford University,\\
Stanford, California 94305, USA}

\address{${}^b$Center for Theoretical Physics and\\
Department of Physics, Massachusetts Institute of Technology,\\
Cambridge, Massachusetts 02139, USA}

\begin{abstract}

We conjecture that the quantum complexity of a holographic state is dual to the action of a certain spacetime region that we call a Wheeler-DeWitt patch. We illustrate and test the conjecture in the context of neutral, charged, and rotating black holes in AdS, as well as black holes perturbed with static shells and with shock waves.   This conjecture evolved from a previous conjecture that complexity is dual to spatial volume, but appears to be a major improvement over the original. In light of our results, we discuss the hypothesis that black holes are the fastest computers in nature.

\end{abstract}

\let\thefootnote\relax\footnotetext{Published in PRL {\bf 116}, 191301 as ``Holographic Complexity Equals Bulk Action?''}

\end{titlepage}

The interior of a black hole is the purest form of emergent space:  once the black hole has formed, the interior grows linearly for an exponentially long time.  
One of the few holographic ideas about the black hole interior is that its growth is dual to the growth of quantum complexity \cite{Cgeodesicduality,Susskind:2014rva}. This duality is a conjecture but it has passed a number of tests.

In the context of AdS/CFT duality, the conjecture has taken a fairly concrete form: the volume of a certain maximal spacelike slice, which extends into the black hole interior, is proportional to the computational complexity of the instantaneous boundary conformal field theory (CFT) state \cite{complexityshocks}. 
The conjecture is an example of the proposed connection between tensor networks and geometry---the geometry being defined by the smallest tensor network that prepares the state. (See also \cite{entrenholo,vidal_tns_geo,hartmanmaldacena, Cgeodesicduality, Susskind:2014rva, localshocks, Susskind:2014jwa}.)

For the case of the two-sided AdS black hole the conjecture is schematically described by
\be
\textrm{Complexity} \sim \frac{V}{G \lads},
\label{CV}
\ee
where  $V$ is the volume of the Einstein-Rosen bridge (ERB), $\lads$ is the radius of curvature of the AdS spacetime, and $G$ is Newton's constant.
Multiplying and dividing   Eq.~\ref{CV}  by $\lads$ suggests a new perspective on the identification of complexity and geometry,
\be
\textrm{Complexity}  \sim \frac{\CW}{G \lads^2},
\ee
where $\CW \equiv \lads V$ now has units of spacetime volume and represents the world volume of the ERB. Further noting that $1/\lads^2$ is proportional to the cosmological constant of the AdS space inspires a new conjecture which we suspect may have deep implications for the connection between quantum information and gravitational dynamics. We propose:
\begin{equation}
\textrm{CA-conjecture:} \ \ \ \ \ \  \textrm{Complexity} = \frac{\textrm{Action}}{\pi \hbar} .  \label{eq:ourconjecture} \hspace{3.5cm}
\end{equation}
(The detailed calculations are presented in \cite{ToAppear}.) The systems we will consider are those whose low-energy bulk physics  is described by the Einstein-Maxwell action
 \begin{equation}
\textrm{Action} =   \frac{1}{16 \pi G} \int_{\mathcal{M}}  \sqrt{|g|} \left( \mathcal{R} - 2 \Lambda  \right) - \frac{1}{16 \pi } \int_{\mathcal{M}}  \sqrt{|g|} F_{\mu \nu} F^{\mu \nu}  +  \frac{1}{8 \pi G} \int_{\partial \mathcal{M}} \sqrt{|h|} {K} , \label{eq:EinsteinHilbertActionForReal}
\end{equation}
with the usual conventions \cite{Poisson}. The three terms in Eq.~\ref{eq:EinsteinHilbertActionForReal} representing the action of a region
$\CM$
 are the Einstein-Hilbert (EH) action including a (negative) cosmological constant, a Maxwell term, and a York-Gibbons-Hawking (YGH) surface term constructed from the extrinsic curvature tensor $K.$

\begin{figure}
\centering
\includegraphics[width=.9\textwidth]{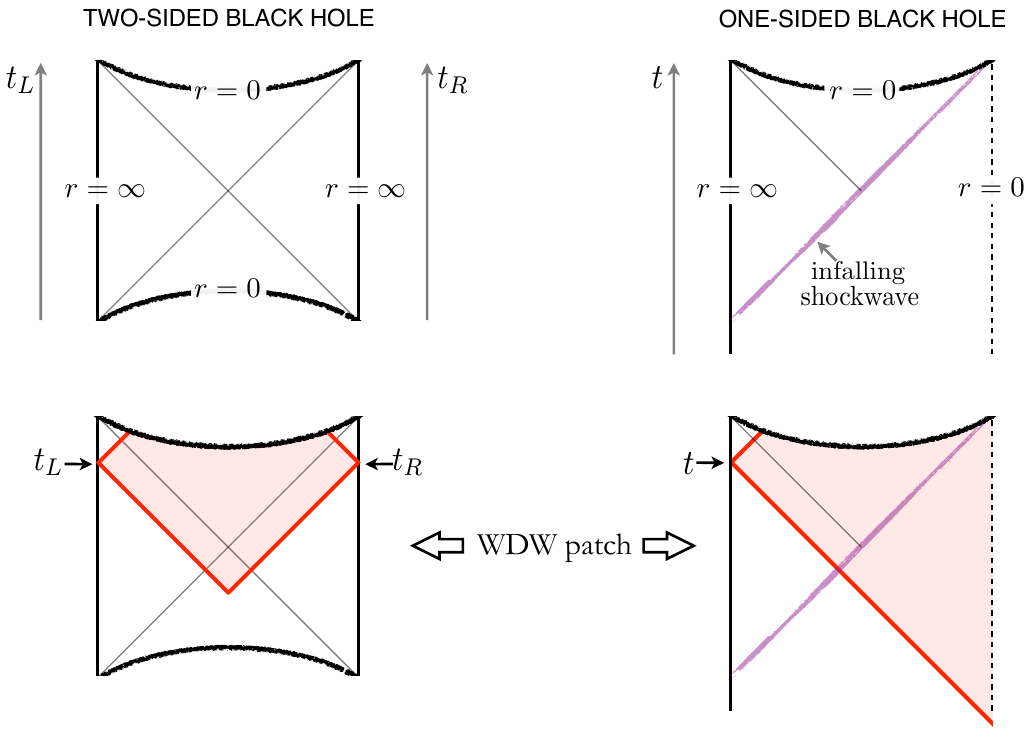}
\caption{The two-sided eternal AdS black hole (left) and a one-sided AdS black hole that forms from a collapsing shockwave (right). The two-sided AdS black hole is dual to an entangled (thermofield double) state of two CFTs that live on the left and right boundaries; the one-sided black hole is dual to a single CFT. Our  complexity/action conjecture relates the complexity of the CFT state to the action of the Wheeler-DeWitt patch (shown shaded).}
\label{fig:summaryfigure}
\end{figure}

In AdS/CFT, the spacetime region  dual to the boundary state is the  ``Wheeler-DeWitt (WDW) patch" whose action, according to the conjecture, gives the complexity of the state. The WDW patch, plotted in Fig.~\ref{fig:summaryfigure},  is given by the union of all spatial slices anchored at a given boundary time $t$ (or pair of times $(t_L, t_R)$ for the two-sided case).

\begin{centering}
\center
$\blacklozenge$$\blacklozenge$$\blacklozenge$$\blacklozenge$
\center
\end{centering}

Black holes are known to excel at information theoretic tasks: they are the densest memory \cite{'tHooft:1993gx,Susskind:1994vu, Bekenstein:1980jp, Marolf:2003sq, Casini:2008cr}; they are the fastest scramblers \cite{bhmirror, fastscramble, Maldacena:2015waa}. Here we explore the possibility that black holes also saturate a universal bound on complexity growth. Computational complexity is the minimum number of quantum gates from some universal set required to prepare the boundary state from a reference state \cite{complexityreview1,complexityreview2}. The reference state may be taken to be an unentangled state so that all non-trivial correlations are accounted for, or simply the initial state if we are interested only in complexity growth \cite{ToAppear}. The use of discrete gates can be justified by lattice regulating the field theory of interest and making a renormalization group argument \cite{ToAppear}.

Interpreting complexity growth as computation, and normalizing complexity using Eq.~\ref{eq:ourconjecture}, we find our proposals are compatible with the saturation of Lloyd's conjectured bound\footnote{The numerical coefficient in Eq.~\ref{eq:Lloyd} is not fixed by our considerations. Indeed the normalization of complexity depends on precise details of the quantum circuits used to prepare the state. What our claim means is that once the normalization is fixed by Eq.~\ref {eq:Lloyd}  for any particular black hole, the same coefficient determines the complexity-action relation for all black holes, and indeed all systems.

We find it interesting that by adopting the  normalization of Lloyd \cite{lloyd2000ultimate} (see also \cite{Margolus:1997ih,PhysRevLett.90.167902}), an increase of complexity by one gate advances the phase of $e^{i \textrm{Action}/\hbar}$ from 1 to -1.} on the rate of computation for a system of energy $M$\cite{lloyd2000ultimate}
\begin{equation}
\frac{d \textrm{Complexity}}{dt} \leq \frac{2 M}{\pi \hbar} . \label{eq:Lloyd}
\end{equation}

The bottom left panel of Fig.~\ref{fig:summaryfigure} shows the WDW patch for a neutral two-sided black hole in AdS. There are two boundary times, one on each side of the wormhole, and the symmetry ensures that the action is only a function of the sum $t_L + t_R$. Calculating the total action of the WDW patch requires a regulator, since  the relevant integrals diverge at the asymptotic AdS boundaries. However, the divergences are independent of time and do not affect the rate of change of  action.

We have found that the late-time rate of change of action  of the WDW patch of the neutral AdS black hole is \cite{ToAppear}
\begin{equation}
\frac{d \textrm{Action}}{d(t_L + t_R)} = 2M. \label{eq:actionincrease}
\end{equation}
This result is simple, but the derivation is non-trivial. It involves a complicated cancellation between EH volume term, and the YGH surface term in  Eq.~\ref{eq:EinsteinHilbertActionForReal}. Remarkably, this result holds for black holes of any size---small, intermediate, or large compared to the AdS radius---and in any number of spacetime dimensions. In the previous proposal of \cite{complexityshocks},  the coefficient in the rate of growth of complexity depended both on the size of the black hole and on the number of dimensions, which made it impossible to saturate a universal bound of the form of  Eq.~\ref{eq:Lloyd}. The universality of the rate of growth of action means that our CA-duality implies that all neutral black holes of any size and in any number of dimensions saturate the same bound with the same coefficient.

Additional evidence for our conjecture is provided by black holes with conserved charge $Q$. With a conserved charge, the system is more restricted and should complexify slower. In \cite{ToAppear} we will argue that the bound should generalize to
\begin{equation}
\frac{d \textrm{Complexity}}{dt} \leq \frac{2}{\pi \hbar} \left[ (M - \mu Q) - (M - \mu Q)_{\textrm{ground state}}\right] .\label{eq:chargedrotatingbound}
\end{equation}
For a given chemical potential $\mu$, the ground state is the state of minimum $M - \mu Q$. No new coefficient is required in this equation---the coefficient is fixed by requiring that Eq.~\ref{eq:chargedrotatingbound} reduce to Eq.~\ref{eq:Lloyd} in the limit $\mu \rightarrow 0$.  For rotating black holes the charge is the angular momentum $J$ and the chemical potential is the angular velocity $\Omega$.

We will now study a number of special cases, choosing the dimensionality to make the calculations easy. Our conclusions should apply in any number of dimensions.

For rotating black holes in 2+1-dimensional AdS, the ground state has $M, J \rightarrow 0$, and we have found that the bound of Eq.~\ref{eq:chargedrotatingbound} becomes \cite{ToAppear}
\begin{equation}
\frac{d \textrm{Complexity}}{dt} \leq \frac{2}{\pi \hbar} (M - \Omega J) = \frac{2}{\pi \hbar} \sqrt{M^2 - \frac{J^2}{\lads^2} }.
\end{equation}
The rate of change of  action of the WDW patch of a rotating black hole in 2+1 dimensions has also been calculated \cite{ToAppear}. Assuming action and complexity are related by our conjecture, Eq.~\ref{eq:ourconjecture}, the bound is again saturated. As in the static case, the action calculation involves nontrivial cancellations between  EH volume term and the YGH surface term.

For  electrically charged black holes in 3+1 dimensional AdS that are much smaller than $\lads$, the minimum of $M - \mu Q$ at fixed $\mu$ is at $M, Q \rightarrow 0$. The bound of Eq.~\ref{eq:chargedrotatingbound} becomes
\begin{equation}
\frac{d \textrm{Complexity}}{dt} \leq \frac{2}{\pi \hbar} (M - \mu Q) = \frac{2}{\pi \hbar} \sqrt{M^2 - \frac{Q^2}{G} }.
\end{equation}
At late times, the rate of change of action of the WDW patch of a small charged black hole in 3+1 dimensions can be calculated \cite{ToAppear} and shown, assuming our conjecture, Eq.~\ref{eq:ourconjecture}, to  precisely saturate this bound. The action calculation involves intricate cancellations, this time between all three terms in Eq.~\ref{eq:EinsteinHilbertActionForReal}.

\begin{figure}
\centering
\includegraphics[width=.4\textwidth]{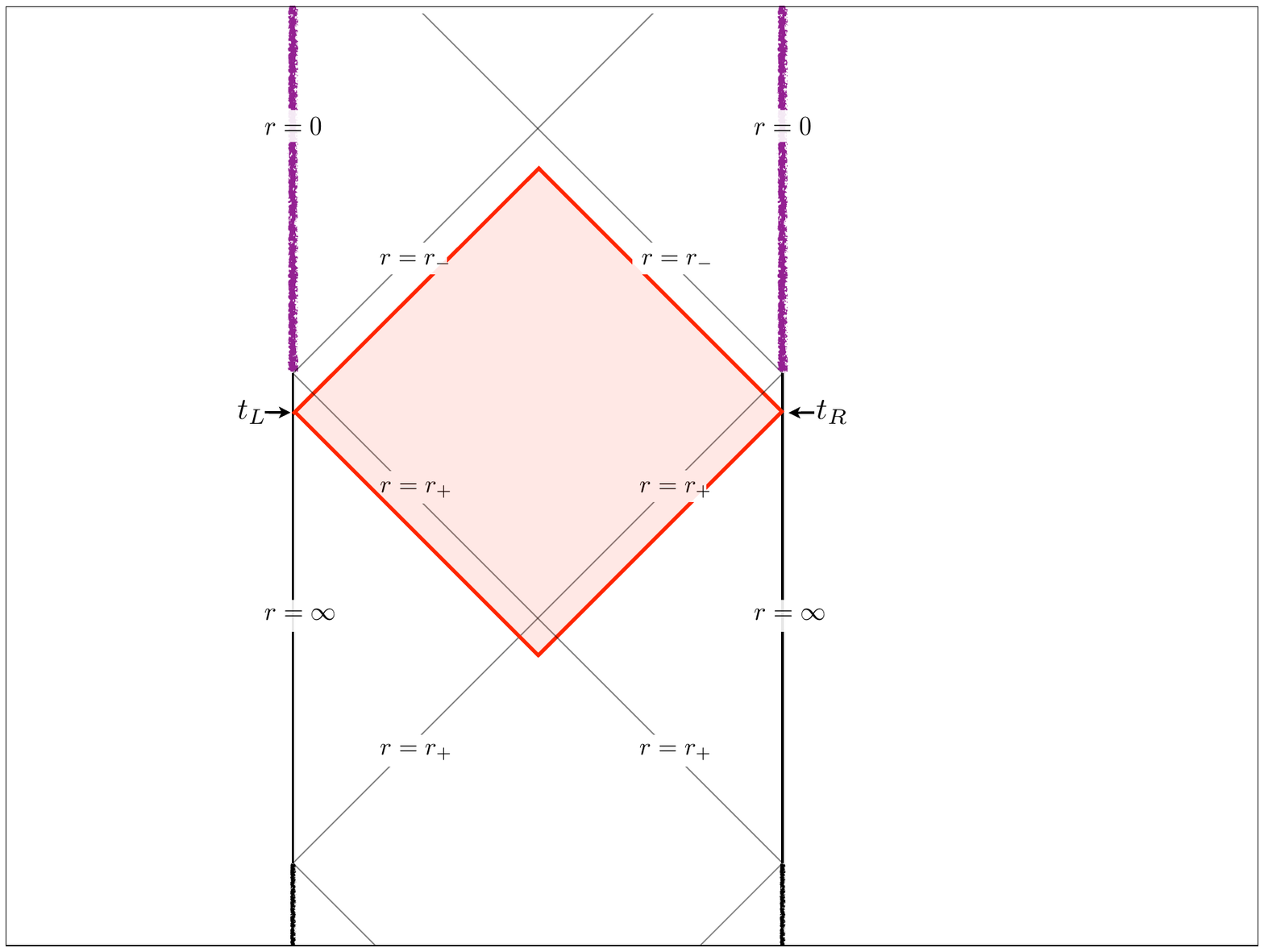}
\caption{For a charged AdS black hole, the Wheeler-DeWitt patch does not extend all the way to the singularity, and instead ends when the ingoing lightsheets self-intersect just outside the inner horizon at $r_-$.}
\label{fig:charged}
\end{figure}

For  electrically charged black holes in 3+1 dimensional AdS that are much larger than $\lads$ (shown in Fig.~\ref{fig:charged}), the situation is more complicated. Large highly-charged black holes have large $\mu$; for large enough $\mu$, the quantity $M - \mu Q$ has a nontrivial minimum less than zero. Naively taking this minimum to be  the large extremal black hole at fixed $\mu$, near extremality the bound of Eq.~\ref{eq:chargedrotatingbound} becomes
\begin{equation}
\frac{d \textrm{Complexity}}{dt} \leq  \frac{4}{\pi \hbar} (M - M_Q) + O\left[ (M-M_Q)^2 \right] ,  \, \ \ \ \ \ \ \ \ \  [\text{naive}]\label{eq:complexitylargecharged}
\end{equation}
where  $M_Q \equiv \frac{2}{3} \left(\frac{G}{3} \right)^{-\frac{1}{4}  }\sqrt{ \frac{Q^3}{\lads}}$. The bound goes linearly to zero at extremality. On the other hand, the late-time rate of change of action of the WDW patch of a large RN black hole in 3+1 dimensions can be calculated \cite{ToAppear} to be
\begin{equation}
\frac{d \textrm{Action}}{dt}  =  \sqrt{6 M_Q(M - M_Q)} + O[M- M_Q]. \label{eq:actionlargeextremal}
\end{equation}
Near extremality this is much larger than Eq.~\ref{eq:complexitylargecharged}, and so  large charged black holes apparently violate the complexity bound. (This problem  also afflicts the complexity-volume conjecture of \cite{complexityshocks}.)

However, the naive bound of Eq.~\ref{eq:complexitylargecharged} would not apply if the extremal RN
black hole were not the state of lowest $M - \mu Q$ at fixed $\mu$. In theories with light charged particles, large RN black holes grow hair. A large enough chemical potential surrounds the black hole with a ball of charged particles, with the ball extending a macroscopic distance from the horizon. This hair has negative $M - \mu Q$; since it ignored this hair, there is no reason to trust the naive calculation deriving Eq.~\ref{eq:complexitylargecharged}.

In order for this resolution to work, all Reissner-Nordstrom-AdS-type large charged black holes that can be embedded in UV-complete theories must develop hair.
%
%
%
Given the absence of a no-hair theorem this does not seem impossible. General arguments have been given in this direction based on the weak-gravity conjecture \cite{weakgravity}, and several explicit examples are known, including superconducting condensation of charged fields \cite{sclandscape,semilocal,sean_review} (such fields are typically present in UV completions of Eq.~\ref{eq:EinsteinHilbertActionForReal}).
Turning this around, the apparent violation of the complexity-bound can be used as a tool to diagnose the development of hair.

This subtlety does not arise in the other cases we have considered. There are good reasons to be believe that neutral, rotating, and small charged AdS black holes can be embedded in UV-complete theories without developing hair. The reason is supersymmetry. There are examples of each of these kinds of black hole in which the black hole is protected from the development of significant hair by the BPS bound. 
The only case for which we expect hair to be inevitable is the one where the naive bound fails.

Taking the example of a superconducting instability, we may ask whether fluctuations about the true superconducting ground state obey the complexity growth bound. The finite temperature solution corresponds to a far-from-extremal charged black hole surrounded by shell of condensed charge \cite{holosc1}. The zero-temperature limit of this solution has all the charge residing in the condensate shell and has vanishing horizon area \cite{holosc2}, and hence does not complexify. This fact along with the known power-law heat capacity of the hairy black hole \cite{holosc1} imply that the complexity growth-bound is qualitatively obeyed (see also the discussion of static shells below).

\begin{centering}
\center
$\blacklozenge$$\blacklozenge$$\blacklozenge$$\blacklozenge$
\center
\end{centering}

A strong test of the relationship between geometry and complexity is provided by perturbing the black hole. In \cite{shock} (see also \cite{Shenker:2013yza,Leichenauer:2014nxa,localshocks,Roberts:2014ifa,Shenker:2014cwa}), the field theory was perturbed with a small thermal-scale operator. In the boundary, this small perturbation grows due to the butterfly effect, and increases the complexity. In the bulk, the perturbation gives rise to an ingoing null shock wave that increases the volume and action of the Einstein-Rosen bridge. Both the complexity-volume duality of \cite{complexityshocks,localshocks} and the complexity-action duality of this paper successfully have these two growths match.

The match is remarkably detailed. Not only does the bulk shock wave calculation successfully reproduce the chaotic growth of complexity in the boundary state, it also reproduces the partial cancellation that occurs during the time it takes the perturbation to spread over the whole system (the ``scrambling time'' \cite{bhmirror, fastscramble}). That the action calculation is sensitive to this cancellation is evidence that it counts the gates of the \emph{minimal} circuit.

The shock wave tests can be made even more stringent. As was shown for the spatial volume of the ERB in \cite{complexityshocks, localshocks}, and as will be shown for its action in  \cite{ToAppear}, we can add more than one shock wave \cite{Shenker:2013yza} in more than one location \cite{localshocks}, and the dual calculations continue to match. (The action calculations that will be presented in \cite{ToAppear} are much easier than the volume calculations of \cite{complexityshocks,localshocks}, since there are now no differential equations to be solved.) These multiple shock wave states provide detailed evidence for the duality between complexity and geometry because the correspondence continues for all possible times and locations of the perturbations.

Another test of the relationship between complexity and geometry is provided by a different kind of perturbation. Rather than sending a null shockwave into the black hole, we will instead surround the black hole with a static shell held aloft by compressive strength. In calculating the rate of change of action, the shell itself does not contribute since it is static. The only effect is indirect---the shell places the black hole in a gravitational well and so gravitational time dilation slows the black hole's rate of action growth. This action calculation fits with our complexity expectation. First, not all energy computes---the static shell is computationally inert. Second, gravitational time dilation makes  computers placed in gravitational wells run slow.

(One way of understanding superconducting black holes is as non-extremal black holes surrounded by a computationally inert superconducting shell.)

\begin{centering}
\center
$\blacklozenge$$\blacklozenge$$\blacklozenge$$\blacklozenge$
\center
\end{centering}

We have introduced a new conjecture in this paper: that the complexity of a holographic state is dual to the action of the associated Wheeler-DeWitt patch. Although motivated by the older complexity/volume duality of \cite{complexityshocks}, the new conjecture subsumes the old. One may ask in what way is it an improvement?

The original conjecture had some ad hoc features, most notably the introduction of a new length scale with each new configuration to be studied. For large black holes in a given AdS background the length scale was chosen to match the AdS radius of curvature. For small AdS black holes or black holes in flat spacetime the scale was the Schwarzschild radius. No such arbitrary scale is needed in the duality relating complexity and action.

The complexity-action duality has been subjected  to a number of non-trivial tests. In \cite{ToAppear} we will subject it to a broader battery of tests, perform more detailed calculations, and discuss the connection with tensor networks. A number of open questions remain. We would like to better understand the implications of the conjecture not just for the rate of change of complexity, as we have in this paper, but for the ground state complexity already present at $t=0$. Finally, as a means of testing our complexity bound, we would like to extend our proposal to less strongly-coupled boundary theories; this will introduce higher-derivative terms in the bulk action, and we will need to be careful to understand their contribution near the singularity.



The coarse-grained geometry of the Einstein-Rosen bridge is given by the circuit (or tensor network) of least complexity that connects the two sides. One wonders if there is a connection between the principle of least action and this principle of least computation.

The complexity-action conjecture relates the geometry of the bulk to the computational complexity of the boundary. The CA-duality provides a tool for diagnosing when horizons are transparent \cite{opacity}, and also for diagnosing {when the state does not belong to a consistent truncation of a UV-complete theory}. The WDW patch is the natural bulk spacetime region to associate with a boundary state, and is robust against small perturbations. The action is a natural quantity associated with the Wheeler-DeWitt patch that  generalizes to higher dimensions, to more general  theories, and to more complicated semiclassical states, without having to make  arbitrary choices. Using CA-duality, we saw that neutral AdS black holes in any number of dimensions and of any size all saturate the same bound on the rate of computation with the same coefficient. (The same coefficient also applies to charged and rotating black holes.)

If our complexity-action conjecture is correct, then  black holes saturate Lloyd's proposed limit on the rate of computation \cite{lloyd2000ultimate}. CA-duality thus provides a natural framework in which to think about black holes as the fastest computers in nature.

\section*{Acknowledgments}
We thank Patrick Hayden, Simon Ross, and Douglas Stanford for useful discussions. AB thanks the participants at Perimeter Institute's ``Quantum Information in Quantum Gravity II'' conference for useful feedback. DR is supported by the Fannie and John Hertz Foundation and is very thankful for the hospitality of the Stanford Institute for Theoretical Physics during various stages of this work. DR also acknowledges the U.S. Department of Energy under cooperative research agreement Contract Number DE-SC00012567. BS is supported by the Simons Foundation. This work was supported in part by National Science Foundation grant 0756174 and by a grant from the John Templeton Foundation.
The opinions expressed in this publication are those of the authors and do not necessarily
reflect the views of the funding agencies.

\bibliographystyle{ut}
\bibliography{SHORTLETTER}

\end{document}